\documentclass[12pt]{article}
\usepackage{epsfig}

\newfont{\Bigtitle}{cmr17 scaled\magstep2}
\newfont{\bigheader}{cmr14}
\newfont{\little}{cmr10}
\newfont{\emt}{cmti10}

\begin{document}
\begin{flushright}
\vspace*{-10mm}
DTP-00/56\\
\tt{hep-ph/0009267}
\end{flushright}
\vspace{5mm} 
\sloppy

\begin{center}


{\LARGE{\bf Is the $\sigma(600)$ a Glueball?}}

\vspace*{1.5mm}

{\LARGE{\bf Two photon reactions can tell us} }
{\footnote{\little Talk given at the 13th International Workshop on Photon-Photon Collisions (Photon2000), September 2000, Ambleside, U.K. --- to appear in the Proceedings}}\\[7mm]
{\bigheader{M.R. Pennington}}\\[5mm]
{\little
{\emt Centre for Particle Theory, University of Durham,
 Durham DH1 3LE, U.K.}}\\[7mm]

\end{center}
{\little
\noindent{\bf Abstract:}
Minkowski and Ochs have recently  argued that the small two photon 
coupling of a
conjectured $\sigma(600)$ is so small that it is likely to be a glueball. 
We ask whether this can be so or whether it is simply gauge invariance that 
produces the observed low mass suppression?}\\[-2mm]

\parskip=1.0mm
\baselineskip=5.8mm

\begin{center}
{\bigheader\bf THE SCALAR GLUEBALL}
\end{center}

QCD predicts that there should exist bound states of glue: states we know as {\it glueballs}~[1]. 
All modellings of non-perturbative QCD, including lattice computations~[2,3],
indicate that the lightest glueball should be spinless. It is then not surprising that if we look at the Particle 
Data tables~[4], there are more light isosinglet scalars  than can fit into one $q{\overline q}$ nonet:
$f_0(400-1200)$ (or $\sigma$), $f_0(980)$, $f_0(1370)$, $f_0(1500)$ and $f_0(1710)$. But which one is the glueball?

Each state has its own protagonists. Lattice calculations expect the scalar glueball has a mass (in the quenched approximation) between
1600 and 1700 MeV~[2,3]. Amsler and Close~[5] claim the $f_0(1500)$,
studied extensively (but not exclusively) in
${\overline p}p$ annihilation with Crystal Barrel at LEAR, 
is the glueball, while others~[3], even more vehemently, insist the scalar component of the $f_J(1710)$~[4] observed in
$J/\psi$ radiative decays is gluish. Here we concentrate on the claim that it is the $\sigma(600)$ that is predominantly a glueball~[6,7].
Though Minkowski and Ochs~[6] are not the first to make such a proposal, it is the appearance of the $\sigma(600)$ in $\gamma\gamma$ reactions that is central to their claim, and so most appropriate for this meeting. 

\begin{center}
{\bigheader \bf TWO PHOTON v. DI-PION PRODUCTION OF PIONS}
\end{center}

In Fig.~1, we compare the cross-section for $\pi^+\pi^-\to\pi^0\pi^0$ from the very recently published results from  
the E852 experiment at Brookhaven~[8] with the Crystal Ball data~[9] on $\gamma\gamma\to\pi^0\pi^0$. Each is dominated by the spin two
$q{\overline q}$ state, the $f_2(1270)$. Consequently, it is at this peak that these cross-sections (or rather the squares of the moduli of the corresponding  amplitudes) are normalised.
Below 1~GeV we see the well-known enhancement
in the $\pi\pi$ elastic cross-section, identified as the $\sigma(600)$, while in the $\gamma\gamma$ reaction
little is seen. It is this smallness around 600 MeV that Minkowski and Ochs~[6] claim points to the $\sigma$ being predominantly glue.

\begin{figure}[t]
\begin{center}
\mbox{~\epsfig{file=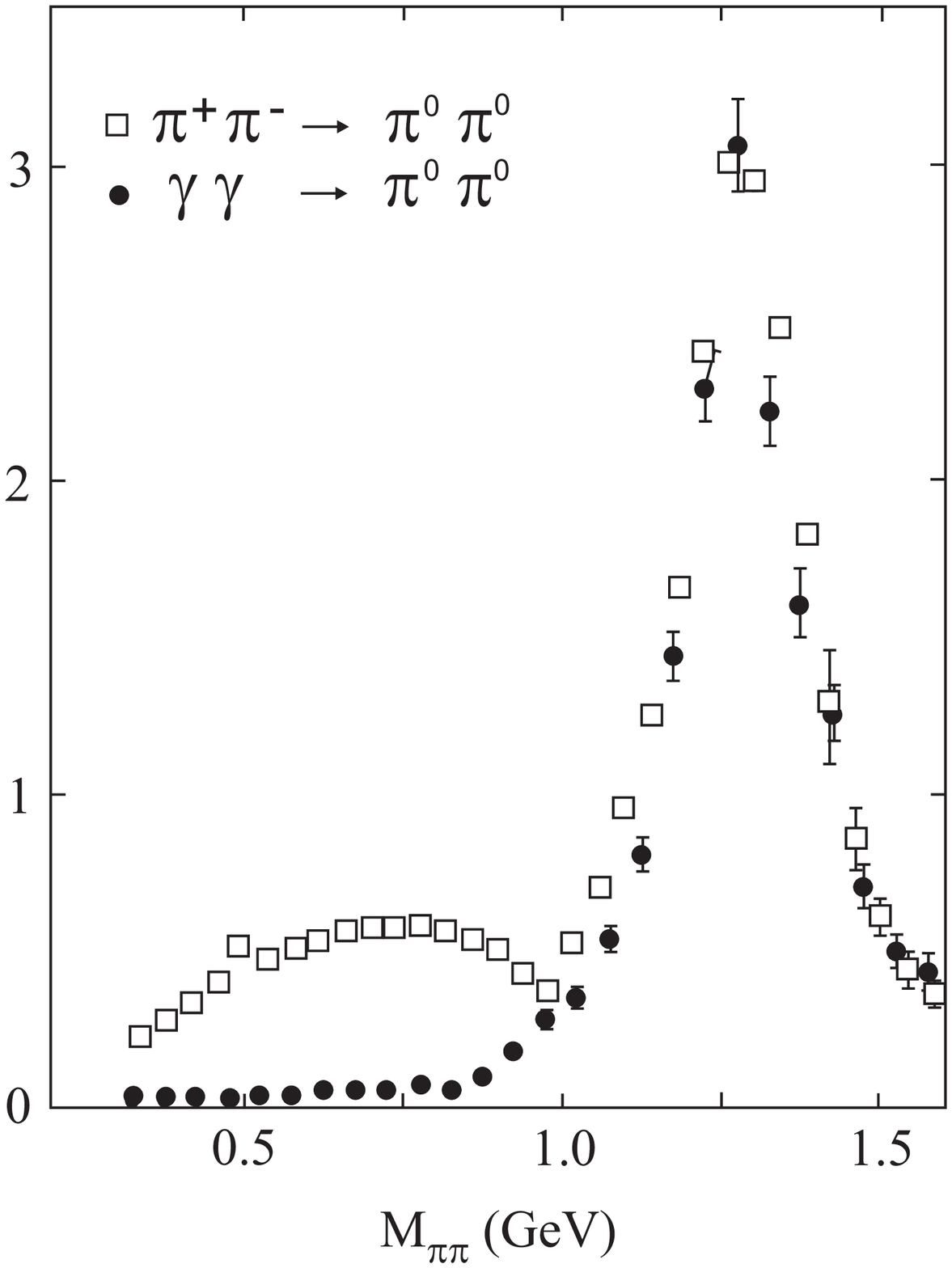,width=8.cm}}
\end{center}
\vspace{-4mm}

\noindent {\bf Figure 1:} {The modulus squared of the amplitudes for $\pi^+\pi^-\to\pi^0\pi^0$ and $\gamma\gamma\to\pi^0\pi^0$, deduced from BNL-E852~ [8] and Crystal Ball~[9] data, respectively.}
\vspace{-3mm}
\end{figure}

In the quenched approximation this is natural. Photons, of course,  couple to electric charge. Consequently,
a neutral $q{\overline q}$ meson can readily decay into photons. In contrast, a glueball can only decay radiatively through
quark loops, Fig.~2. This means that, in the quenched approximation, a glueball 
doesn't have photonic decays. In perturbative QCD, quark loops (Fig.~2) are suppressed by factors of $\alpha_s$.
 But are these arguments relevant at 600 MeV?
Indeed, if the $\sigma(600)$ were to be a glueball, it would have to have large couplings to quarks, if its decay width is to be 300-500 MeV as experiment indicates (Fig.~1).

\begin{figure}[b]
\vspace{-1mm}
\begin{center}
\mbox{~\epsfig{file=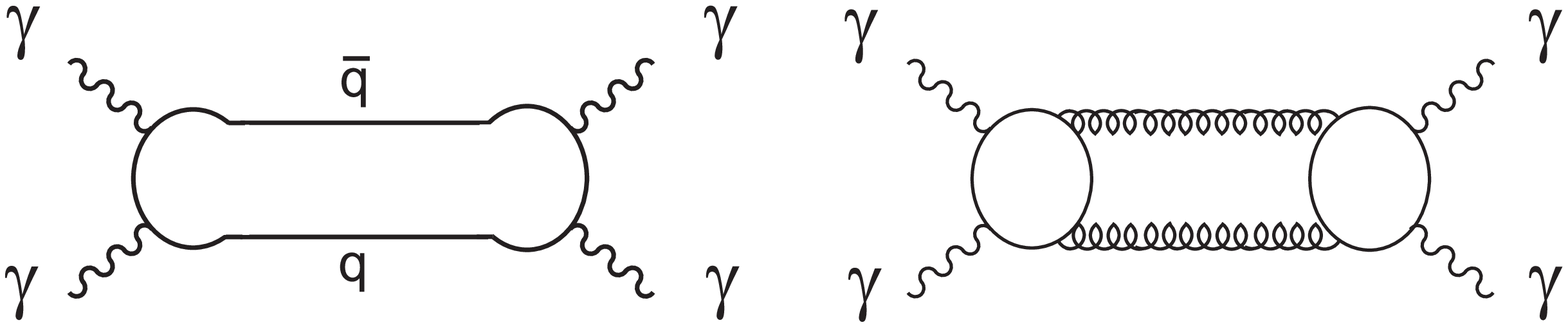,width=12.8cm}}
\end{center}

\noindent {\bf Figure 2:} {Modulus squared of the amplitude for a ${\overline q}q$ state and a glueball, respectively, to decay to two photons
 --- at lowest order in perturbative QCD.}
\vspace{-3mm}
\end{figure}

\newpage
\baselineskip=6mm

\begin{figure}[h]
\vspace{1mm}
\begin{center}
\mbox{~\epsfig{file=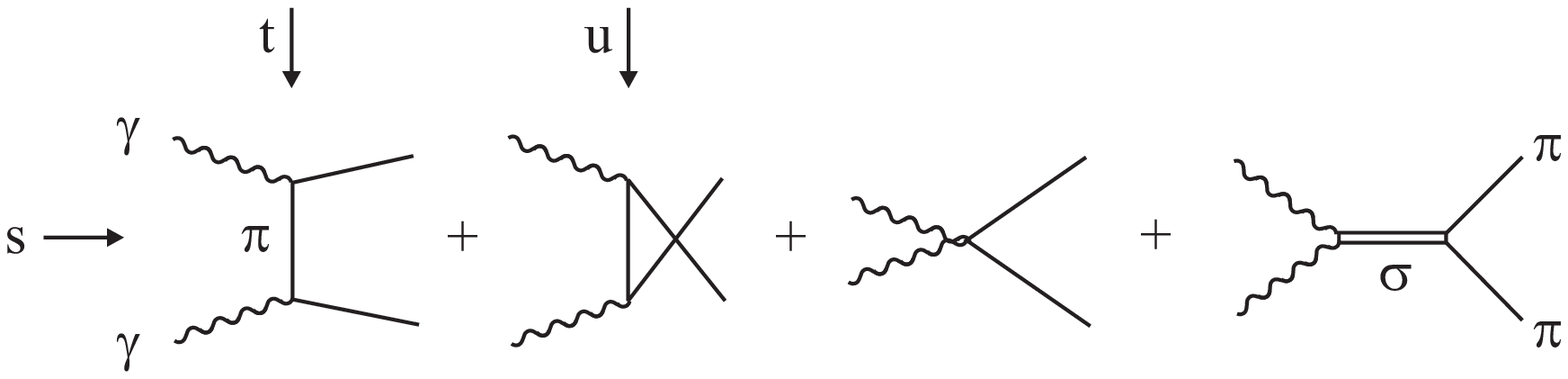,height=3.4cm}}
\end{center}

\noindent{\bf Figure 3:} {Feynman diagram modelling of $\gamma\gamma\to\pi\pi$: the Born term plus direct channel $\sigma$ formation.}
\end{figure}

The idea that the $\gamma\gamma\to\pi^0\pi^0$ cross-section directly determines the $\sigma$ couplings to two photons
requires a modelling of the $\gamma\gamma$ reaction. The simplest  would be to imagine that the two photon amplitude is
just described by the pion Born term plus the $\sigma-$resonance component, as in Fig.~3. Since the Born term does not contribute to the $\pi^0\pi^0$ channel, the cross-section in Fig.~1 then fixes the magnitude of this $\sigma-$component. But if this model makes any sense it must also explain  the $\pi^+\pi^-$ cross-section~[10] too. Its $S-$wave component~[11], then determines the orientation of this
$\sigma-$component. Moreover, unitarity requires that the  phase of each $\gamma\gamma\to\pi\pi$ partial wave amplitude with definite isospin
must equal the phase of the corresponding $\pi\pi$ elastic partial wave. As detailed in [12], this also fixes the orientation
of the $\sigma-$component in the model of Fig.~3. These determinations don't agree. This teaches us that the modelling shown in Fig.~3 is far too simplistic.
There are in fact many other contributions that must be included, like those in Fig.~4, before we can  be sure that unitarity is fulfilled
and so extract resonance couplings in a meaningful way.

\begin{figure}[h]
\begin{center}
\mbox{~\epsfig{file=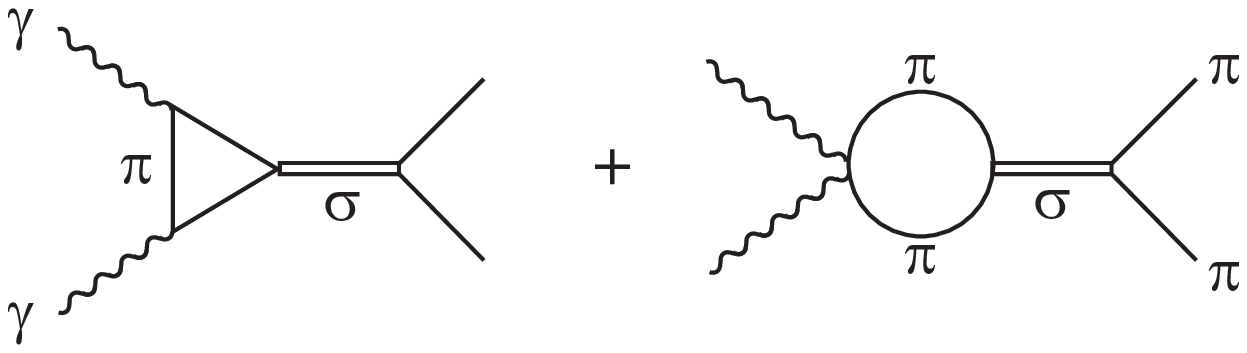,height=2.8cm}}
\end{center}
\vspace{-3mm}

\noindent{\bf Figure 4:} {Two of the additional contributions to the model amplitudes for $\gamma\gamma\to\pi\pi$ of Fig.~3 essential for ensuring the final state interaction theorem is satisfied.}
\end{figure}

\begin{figure}[h]
\begin{center}
\mbox{~\epsfig{file=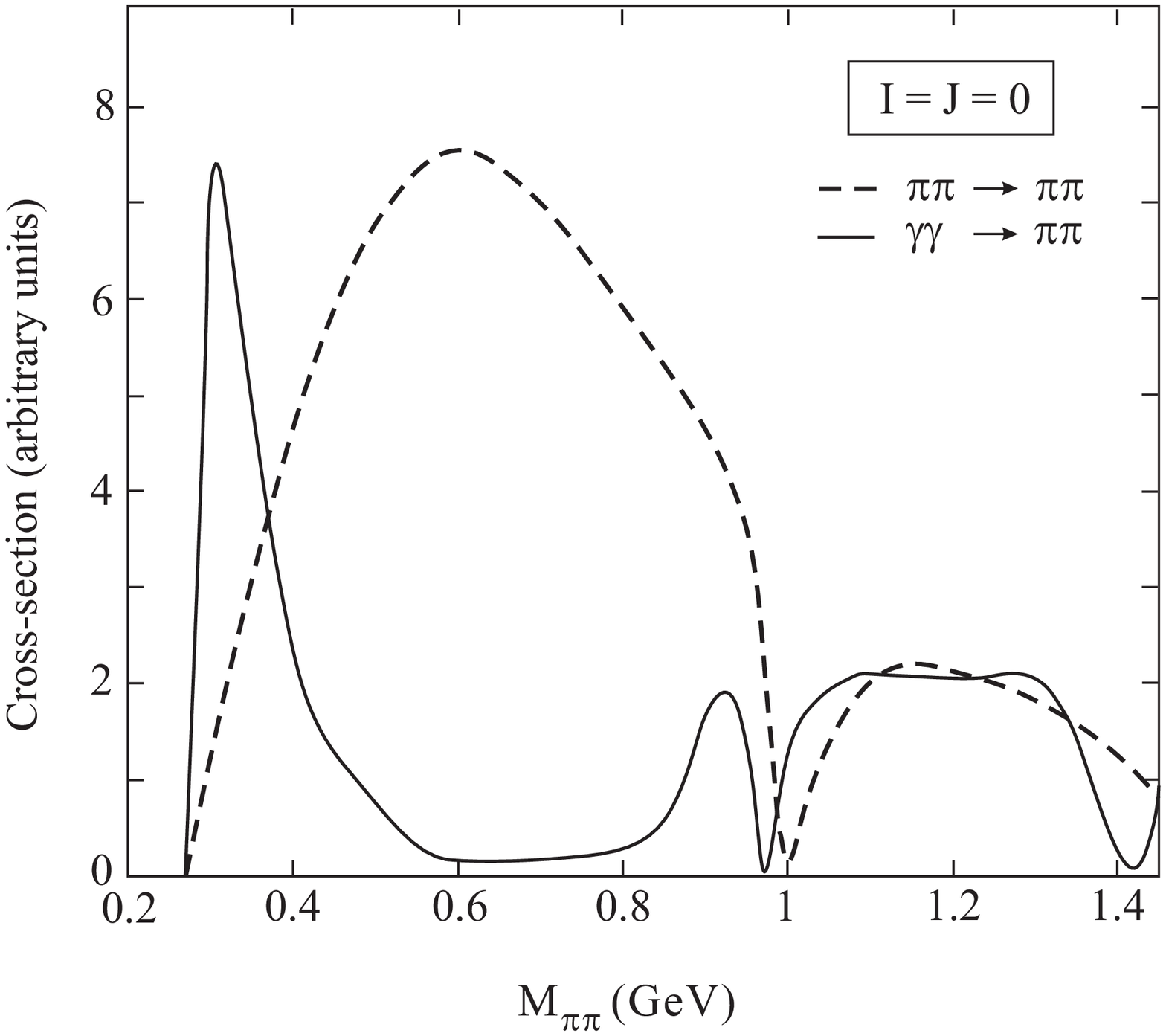,width=10.5cm}}
\end{center}
\vspace{-3mm}

\noindent{\bf Figure 5:} {The $I=J=0$ components of the cross-sections for $\pi\pi\to\pi\pi$ (dashed line) based on data analysed by Ochs~[15] and by Bugg {\it et al.}~[16],
and for $\gamma\gamma\to\pi\pi$ (solid line) from the Amplitude Analysis by Boglione {\it et al.}~[11] --- {\it dip} solution.
}
\end{figure}

Over ten years ago,  David Morgan and I~[13] worked out a procedure for ensuring that $\gamma\gamma\to\pi\pi$ amplitudes satisfy the constraints of analyticity and unitarity,
as well obeying the Thomson limit of QED  at low energies~[14]. Applying this to all recent experimental data, Elena Boglione and I~[11] have found that the $I=0$ $\gamma\gamma\to\pi\pi$ $S-$wave cross-section has a quite different shape than that for $\pi\pi$ scattering, as shown in Fig.~4.
Indeed, while the Born component dominates  the $\gamma\gamma$ channel close to threshold, we see that any resonance contribution is pushed to higher mass away from 600 MeV. There must be some reason for this.


\begin{center}
{\bigheader \bf GAUGE INVARIANCE}
\end{center}

The simplest Lagrangian describing a scalar field $\phi$ coupling to two photons is
$${\cal L}_I\;=\; g\,\phi\,{\cal F}^{\mu\nu}\,{\cal F}_{\mu\nu}\quad,$$
where ${\cal F}^{\mu\nu}$ is the electromagnetic tensor. The scalar's radiative decay rate is proportional to $g^2$. Dimensional analysis requires $g$ to
have dimension of $1/M$. So $g^2$ must be multiplied by a mass cubed to give the radiative width.
Only a detailed non-perturbative calculation would tell us exactly what this mass scale is, but simple phenomenology indicates it must the mass of the scalar
\footnote{\little{Just as for the Higgs boson.}}.
This intuition is confirmed in the pseudoscalar sector. The dimensional analysis  is the same~[17]. In the  quark model, the $\gamma\gamma$ couplings are
related to the fourth power of the charges of the constituents (Fig.~2), but for the $\pi$, $\eta$ and $\eta'$ this has to be corrected by a factor of
their mass cubed to bring agreement with experiment. Without this factor, the quark model predictions would be orders of magnitude wrong
\footnote{\little{Arguments from non-relativistic quark models would give other mass dependent factors too. While such models are applicable to heavy quark systems, it is not clear that they have much relevance to light quark systems, where, as stressed in [18], for instance, genuine bound state calculations, which are inevitably non-perturbative, are essential for meaningful results.}}.

\begin{figure}[t]
\begin{center}
\mbox{~\epsfig{file=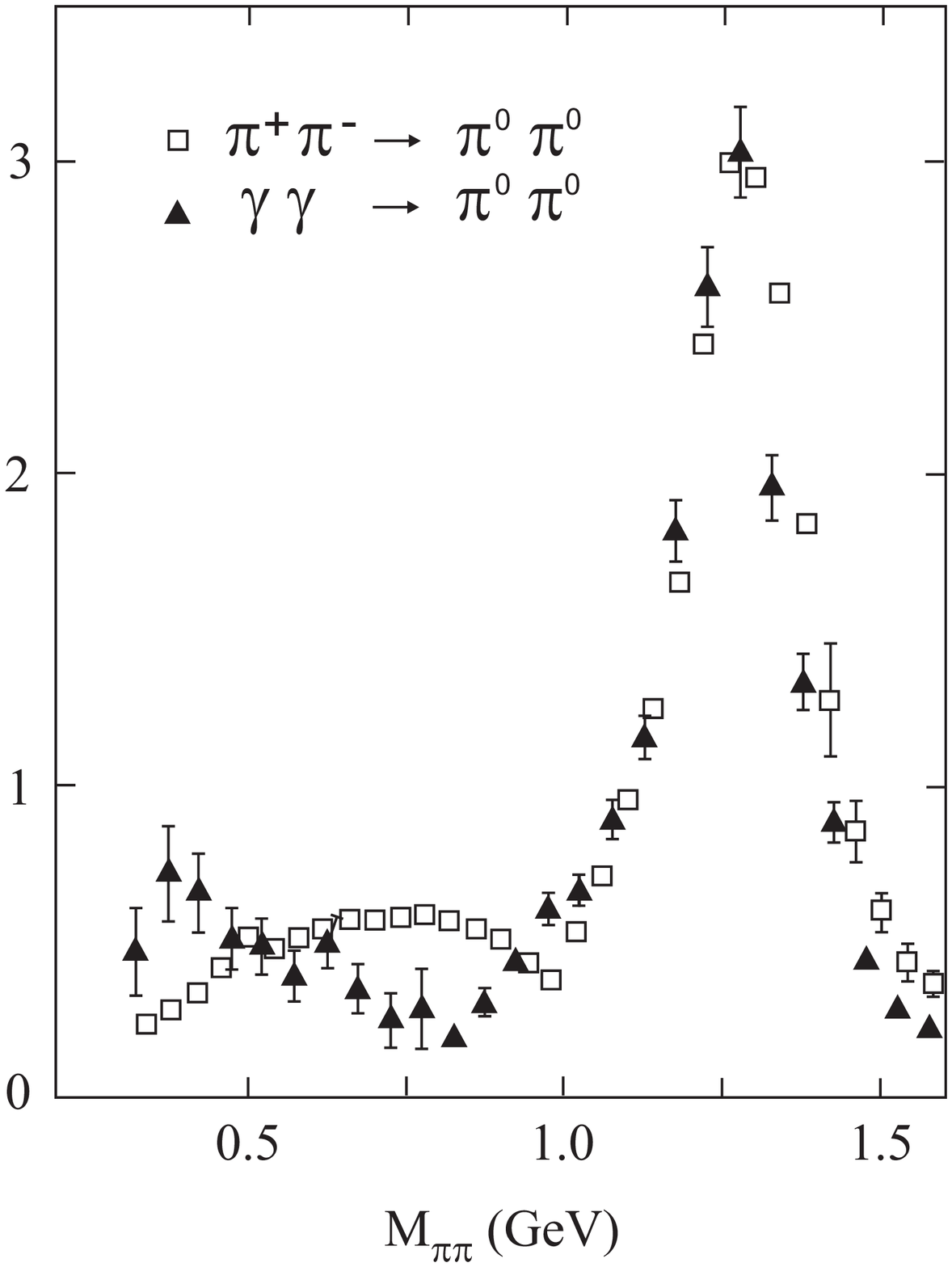,width=8.cm}}
\end{center}
\vspace{-5mm}

\noindent{\bf Figure 6:} {The modulus squared of the amplitudes for $\pi^+\pi^-\to\pi^0\pi^0$ and $\gamma\gamma\to\pi^0\pi^0$, (data as in Fig.~1). Here the $\gamma\gamma$ results are scaled by a factor of $M_{\pi\pi}^{\,-3}$ normalized to 1 at 1270 MeV to be compared with Fig.~1.}
\vspace{-3mm}
\end{figure}

It is then natural to compare the cross-sections of Fig.~1 with the $\gamma\gamma\to\pi^0\pi^0$ one divided by a factor of $M_{\pi\pi}^{\ 3}$, normalised to unity at the $f_2(1270)$. The result is displayed in Fig.~6.
We see that the $\pi\pi$ and $\gamma\gamma$ cross-sections are now much more equal across the whole mass region below 1.5~GeV with no further suppression of the $\sigma$ at 600~MeV.
Clearly the factor of mass cubed cannot be exact as $M_{\pi\pi}$ increases
indefinitely. However, the phenomenology in the pseudoscalar sector indicates that the factor is indeed mass cubed, at least up to the $\eta'$ around 1 GeV, which is sufficient for this illustration.

We see that it is wholly a consequence of  gauge invariance that the $\sigma$
contribution is suppressed at 600 MeV, just as the  radiative width of the pion is reduced compared to the $\eta$ and $\eta'$. 
The $I=0$ $S-$wave $\gamma\gamma\to\pi\pi$ cross-section is skewed to higher masses.  The $f_0(400-1200)$ seen in Fig.~5 has a radiative width of about 4 keV, as discussed in [11]. This is just as expected for a scalar composed of non-strange quarks, i.e. a $(u{\overline u} + d{\overline d})$ state~[19].
Thus the $\sigma(600)$ is not, as claimed by Minkowski and Ochs~[6], the glueball.
For this, we have to look above 1.4 GeV. 

It is an important feature of the scalar sector and its relation to the vacuum that the glueball lies amongst the
lightest ${\overline q}q$ isoscalars and inevitably mixes with these. In contrast, lattice calculators~[1] compute the tensor glueball mass is above 2 GeV and so far above the lightest quark nonet, containing the $f_2(1270)$.
Consequently, any mixing is likely to be less (ignoring the effect of radial excitations) and the glueball nearly naked~[1]. While the $\xi(2230)$~[4,20]
may be a candidate for such a state, it is essential to confirm its existence
and then to determine it does have spin 2, before speculating further about its composition~[21]. 

QCD predicts that bound states of glue should exist.
The scalar is the lightest. Surely soon we will know which of the many 
$f_0$'s~[4] it is.
Two photon reactions tell us it is not the $\sigma(600)$.

\vspace{1mm}
\begin{center}
{\bigheader \bf ACKNOWLEDGEMENTS}
\vspace{-1mm}
\end{center}
It is a pleasure to thank the organisers, particularly
Alex Finch, for this de-lightful meeting in such a wonderful setting.
I acknowledge support from the EU-TMR Programme, Contract No.
CT98-0169, EuroDA$\Phi$NE. 

\vspace{1mm}
\begin{center}
{\bigheader \bf REFERENCES}
\vspace{-1mm}

\end{center}
\baselineskip=5mm
\parskip=0.6mm

{\little
\noindent 1. see, for instance, Pennington, M.R., {\emt Proc. Workshop on
 Photon Interactions and the Photon Structure} (Lund,
September 1998), eds G. Jarlskog, T. Sj\"ostrand, pub. Lund Univ, pp. 312-328. 

\noindent 2. Teper, M., 
{\emt Proc. Int. Europhysics Conf. on High-Energy Physics (HEP 97)} (Jerusalem, Israel, August 1997)  pp. 384-387;  {\lq\lq Glueball masses and other physical properties of SU(N) gauge theories in D=(3+1): A review of lattice results for theorists'', hep-th/981217.

\noindent 3. Sexton, J., Vaccarino, A., and Weingarten, D., {\emt Phys.~Rev.~Lett.} 75, 4563-4566 (1995);

\noindent Lee, W., and Weingarten, D., {\emt Nucl.~Phys.~B (Proc.~Suppl.)} 53, 236-238 (1997), 63, 194-196 (1998);

\noindent Morningstar, C., and Peardon, M., {\emt Nucl.~Phys.~B (Proc.~Suppl.)} 53, 917-920 (1997).

\noindent 4. Groom, D.E., et al. (PDG) {\emt Eur.~Phys.~J.} C15, 1 (2000).

\noindent 5. Amsler, C., and Close, F.E., {\emt Phys.~Lett.} B353, 385-390 (1995),
{\emt Phys. Rev.} D53, 295-311 (1996).

\noindent 6. Minkowski, P., and Ochs, W., hep-ph/9811518, 
{\emt Eur. Phys. J.} C9, 283-312 (1999).

\noindent 7. Kisslinger, L.S., Gardner, J., and Vanderstraeten, C., {\emt Phys.~Lett.} B410, 1-5 (1997);

\noindent Kisslinger, L.S., and Li, Z., {\emt Phys.~Lett.} B445, 271-273 (1999);

\noindent see also, for instance,
Narison, S., {\emt  Nucl.~Phys.} B509, 312-356 (1998).

\noindent 8. Gunter, J., et al., (E852 Collab.) hep-ex/0001038.

\noindent 9. Marsiske, H., et al., {\emt Phys. Rev.} D41, 3324-3335 (1990);

\noindent Bienlein, J.K., {\emt Proc. IXth~Int.~Workshop on Photon-Photon
Collisions} (San Diego, 1992) eds. D. Caldwell and H.P. Paar, pub. World Scientific, pp.~241-257. 

\noindent 10. Boyer, J., et al. (MarkII), {\emt Phys.~Rev.} D42, 1350-1367 (1990);

\noindent Behrend, H.J., et al. (CELLO), {\emt Z.~Phys.} C56, 381-390 (1992).

\noindent 11. Boglione, M., and Pennington, M.R.,  
{\emt Eur.~Phys.~J.} C9, 11-29 (1999).

\noindent 12. Pennington, M.R., {\emt Proc. Workshop on Hadron Spectroscopy}
 (Frascati, March 1999), eds T. Bressani, A. Feliciello and A. Filippi, pub. INFN, 1999, pp. 95-114.

\noindent 13. Morgan, D., and Pennington, M.R., {\emt Z.~Phys.} C37, 431-447 (1988); C39, 590 (1988); C48, 623-632 (1990).

\noindent 14. Pennington, M.R., {\emt DA$\Phi$NE Physics Handbook}, eds. L. Maiani, G. Pancheri and N. Paver, pub. INFN, Frascati, 1992, pp. 379-418;
{\emt Second DA$\Phi$NE Physics Handbook}, eds. L. Maiani, G. Pancheri and N. Paver, pub. INFN, Frascati, 1995,  pp. 531-558.

\noindent 15. Ochs, W., thesis submitted to the University of Munich (1974).

\noindent 16. Bugg, D.V., Zou, B.S., and Sarantsev, A.V., {\emt Nucl.~Phys.} B471, 59-89 (1996).

\noindent 17. Hayne, C., and Isgur, N., {\emt Phys.~Rev.} D25, 1944-1950 (1982);

\noindent Schuler, G.A., Berends, F.A., and van Gulik, R.,, {\emt Nucl.~Phys.} B523, 423-438 (1998).

\noindent 18. Pennington, M.R., {\emt Proc. Int. Conf. on the Structure and Interactions of the Photon} (Freiburg,  May 1999), ed. S. S\"oldner-Rembold, {\emt Nucl.~Phys.~B (Proc.~Suppl.)} 82, 291-299 (2000). 

\noindent 19. Li, Z.P., Close, F.E., and Barnes, T., {\emt Phys.~Rev.} D43, 2161-2170 (1991). 

\noindent 20. Baltrusaitis, R.M. et al. (Mark III), {\emt Phys~ Rev.~Lett.} 56, 107-110 (1986);

\noindent Bai, J.Z., et al. (BES), {\emt Phys.~Rev.~Lett.} 76, 3502-3505 (1996).

\noindent 21.  Chao, K.T., {\emt Commun.~Theor.~Phys.} 24, 373-376 (1995);

\noindent Huang, T., et al., {\emt Phys.~Lett.}  B380, 189-192 (1996);
\noindent Paar, H.P., {\emt Nucl.~Phys.~B (Proc.~Suppl.)} 82, 337-343 (2000).

}

\end{document}